%%%%%%%%%%%%%%%%%%%%%%%%%%%%%%%%%%%%%%%%%%%%%%%%%%%%%%%%%%%%%%%%%%%%%%%%%%%
% Title:  Matrix integrals as Borel sums of Schur function expansions 
% Type:   Lectures by J. Harnad  
%         Symmetry and perturbation theory (SPT 2002) 
%         Cala Ganone, Sardinia, May 19-26, 2002.
%         Nonlinear evolution equations and dynamical systems (NEEDS)
%         Cadiz, Spain, June 9-16, 2002.       
%=============================================================
% Author:    J. Harnad and A. Yu. Orlov
% Compiler:  Plain Tex
% Date:     Sept. 4,   2002
% ===========================================================
%%%%%%%%%%%%%%%%%%%%%%%%%%%%%%%%%%%%%%%%%%%%%%%%%%%%%%%%%%%%%%%%%%%%%%%%%%%%
%
% (Macros in Plain TeX by Aurel Wisse, Version 1.4)
%
%%%%%%%%%%%%%%%%%%%%%%%%%%%%%% Useful Definitions %%%%%%%%%%%%%%%%%%%%%%%%%
\def\@{{\char'100}}

\long\def\abstract#1{\bigskip{\advance\leftskip by 2true cm
\advance\rightskip by 2true cm\eightpoint\centerline{\bf
Abstract}\everymath{\scriptstyle}\vskip10pt\vbox{#1}}\bigskip}
\long\def\resume#1{{\advance\leftskip by 2true cm
\advance\rightskip by 2true cm\eightpoint\centerline{\bf
R\'esum\'e}\everymath{\scriptstyle}\vskip10pt \vbox{#1}}}

\def\references{\bigbreak\centerline{\sc
References}\medskip\nobreak\bgroup
\def\ref##1&{\leavevmode\hangindent45pt
\hbox to 42pt{\hss\bf[##1]\ }\ignorespaces}
\parindent=0pt
\everypar={\ref}\par}
\def\endreferences{\egroup}
\long\def\authoraddr#1{\medskip{\baselineskip9pt\let\\=\cr
\halign{\line{\hfil{\Addressfont##}\hfil}\crcr#1\crcr}}}
\def\Subtitle#1{\medbreak\noindent{\Subtitlefont#1.} }
%
% Runningheads
%
\newif\ifrunningheads
\runningheadstrue \immediate\write16{- Page headers}
\headline={\ifrunningheads\ifnum\pageno=1\hfil\else\ifodd\pageno\rightheadline
\else\leftheadline\fi\fi\else\hfil\fi}
\def\rightheadline{\sc\hfil\RightHeadText\hfil}
\def\leftheadline{\sc\hfil\LeftHeadText\hfil}

\hyphenation{Harnad Neumann}
%
%%%%%%%%%%%%%%%%%%%%%%%%%%%%%%%% Fonts %%%%%%%%%%%%%%%%%%%%%%%%%%%%%%%%%%%%%
%
\immediate\write16{- Fonts "Small Caps" and "EulerFraktur"}
%
%
%%  The following two lines introduce the Small Caps font at 10pt,
%%  if not available comment out and replace by the
%%  following line:
% \def\sc{\rm}
%

\let\sc=\tensmc
%
%%  The following four lines introduce the Euler Fraktur font,
%%  if not available comment out and replace by
%%  the line:
% \def\gr{}
%
\font\teneuf=eufm10  \font\seveneuf=eufm7 \font\fiveeuf=eufm5
\newfam\euffam 

\textfont\euffam=\teneuf \scriptfont\euffam=\seveneuf
\scriptscriptfont\euffam=\fiveeuf
%
%%%%%%%%%%%%%%%%%%%%%%%%%%%% Definitions %%%%%%%%%%%%%%%%%%%%%%%%%%%%%%%%

\font\eightrm=cmr8
\font\eightbf=cmbx8
\font\eightit=cmti8
 
\def\cite#1{{[#1]}}
\def \smalltype{\let\rm=\eightrm  \let\bf=\eightbf
\let\it=\eightit \let\sl=\eightsl \let\mus=\eightmus
\baselineskip = 9.5pt minus .75pt  \rm}

\def \wt {\widetilde}

\def \d {\delta}

\def \G {\Gamma}

\def \l {\lambda}

\def \m  {\mu}

\def \di {\partial}

\def \det{{\rm det}}

\def\nchi{\hbox{\raise 2.5pt\hbox{$\chi$}}}
%
%Bold roman letters
%

 \def\bfF{{\bf F}}

 \def\bfN{{\bf N}}

\def\bft{{\bf t}} 
\def\bftt{{\bf t^*}}

\def\bfx{{\bf x}} 
\def\bfy{{\bf y}} 
 \def\bfZ{{\bf Z}}

%
% Title, Author, Address and Abstract fonts
%
\def\authorfont{\sc} 
\font\eightsl=cmsl8 
\def\eightpoint{\let\rm=\eightrm \let\bf=\eightbf \let\it=\eightit
\let\sl=\eightsl \baselineskip = 9.5pt minus .75pt  \rm}

\font\titlefont=cmbx10 scaled\magstep2 \font\sectionfont=cmbx10
\font\Subtitlefont=cmbxsl10 \font\Addressfont=cmsl8
%
% Section headings
%
\immediate\write16{- Section headings}
\newcount\secount
\secount=0
\newcount\eqcount
\outer\def\section#1.#2\par{\global\eqcount=0\bigbreak \ifcat#10
 \secount=#1\noindent{\sectionfont#1. #2}
\else
 \advance\secount by 1\noindent{\sectionfont\number\secount. #2}
\fi\par\nobreak\medskip}
%
% Definition of automatic numbering
% and corresponding commands
%
\immediate\write16{- Automatic numbering} \catcode`\@=11
\def\adv@nce{\global\advance\eqcount by 1}
\def\unadv@nce{\global\advance\eqcount by -1}
\def\nextnumber{\adv@nce}
%
% Automatic numbering
%
\newif\iflines
\newif\ifm@resection
\def\onesec{\m@resectionfalse}
\def\moresec{\m@resectiontrue}
\moresec
\def\eq{\global\linesfalse\eq@}
\def\eqn{\global\linestrue&\eq@}
\def\nosubind@x{\global\subind@xfalse}
\def\newsubind@x{\ifsubind@x\unadv@nce\else\global\subind@xtrue\fi}
\newif\ifsubind@x
\def\eq@#1.#2.{\adv@nce
 \if\relax#2\relax
  \edef\loc@lnumber{\ifm@resection\number\secount.\fi
  \number\eqcount}
  \nosubind@x
 \else
  \newsubind@x
  \edef\loc@lnumber{\ifm@resection\number\secount.\fi
  \number\eqcount#2}
 \fi
 \if\relax#1\relax
 \else
  \expandafter\xdef\csname #1@\endcsname{{\rm(\loc@lnumber)}}
  \expandafter
  \gdef\csname #1\endcsname##1{\csname #1@\endcsname
  \ifcat##1a\relax\space
  \else
   \ifcat\noexpand##1\noexpand\relax\space
   \else
    \ifx##1$\space
    \else
     \if##1(\space
     \fi
    \fi
   \fi
  \fi##1}\relax
 \fi
 \eq@@{\loc@lnumber}}
\def\eq@@#1{\iflines \else \eqno\fi{\rm(#1)}}
\def\m@th{\mathsurround=0pt}
%
% Command \display
%
\def\display#1{\null\,\vcenter{\openup1\jot
\m@th \ialign{\strut\hfil$\displaystyle{##}$\hfil\crcr#1\crcr}}
\,}
\newif\ifdt@p
\def\@lign{\tabskip=0pt\everycr={}}
\def\displ@y{\global\dt@ptrue \openup1 \jot \m@th
 \everycr{\noalign{\ifdt@p \global\dt@pfalse
  \vskip-\lineskiplimit \vskip\normallineskiplimit
  \else \penalty\interdisplaylinepenalty \fi}}}
%
% Command \displayno
%
\def\displayno#1{\displ@y \tabskip=\centering
 \halign to\displaywidth{\hfil$
\@lign\displaystyle{##}$\hfil\tabskip=\centering&
\hfil{$\@lign##$}\tabskip=0pt\crcr#1\crcr}}
%
% References
%
\def\cite#1{{[#1]}}
\catcode`\@=\active
%
%%%%%%%%%%%%%%%%%%%%%%%%%%%%%%%% Formatting %%%%%%%%%%%%%%%%%%%%%%%%%%%%%%%
%
\magnification=\magstep1 \hsize= 6.75 true in \vsize= 8.75 true in
%
%%%%%%%%%%%%%%%%%%%%%%%%%%%%%%%%%%% Headers %%%%%%%%%%%%%%%%%%%%%%%%%%%%%%
%
\def\RightHeadText{Matrix integrals as Borel sums of Schur function
expansions }
\def\LeftHeadText{J. Harnad and A. Yu. Orlov}
%
%%%%%%%%%%%%%%%%%%%%%%%%% Document %%%%%%%%%%%%%%%%%%%%%%%%%%%%%%%%%%%%%%%%
%
%\rightline{CRM-xxxx (2002) \break} 
%\rightline{solv-int/0209xx \break} 
\bigskip \centerline{\titlefont Matrix integrals as Borel sums of}
\centerline{\titlefont Schur function expansions
\footnote{$^\dagger$}{\smalltype Based on a talk given by J. Harnad at the
workshop: {\it Symmetry and Perturbation theory 2002},  Sardinia (Cala
Gonone),  May 19-26, 2002.}}

\bigskip
\centerline{\authorfont J.~Harnad}
\authoraddr
{Department of Mathematics and Statistics, Concordia University\\
7141 Sherbrooke W., Montr\'eal, Qu\'ebec, Canada H4B 1R6, {\rm
\eightpoint
and} \\
Centre de recherches math\'ematiques, Universit\'e de Montr\'eal\\
C.~P.~6128, succ. centre ville, Montr\'eal, Qu\'ebec, Canada H3C 3J7\\
{\rm \eightpoint e-mail}: harnad\@crm.umontreal.ca}
\bigskip
\centerline{\authorfont  A. Yu. Orlov}
\authoraddr
{Nonlinear Wave Processes Laboratory, \\
Oceanology Institute, 36 Nakhimovskii Prospect\\
Moscow 117851, Russia
\\
{\rm \eightpoint e-mail}: orlovs\@wave.sio.rssi.ru }
\bigskip
\abstract{The partition function for unitary two matrix models is known
to be a double KP $\tau$--function, as well as providing solutions to the
$2$--dimensional Toda hierarchy. It is shown how it may also be viewed as a
Borel sum regularization of divergent sums over products of Schur
functions in the two sequences of associated KP flow variables.}
\bigskip
\baselineskip 14 pt

%%%%%%%%%%%%%%%%%%%%%%%%%%%% Section 1. Introduction %%%%%%%%%%%%%%%%%%%
\section 1. Introduction. $\tau$--functions related to Schur functions. 
\smallskip \nobreak

We recall here a method for constructing a large class of double KP 
$\tau$--functions in terms of infinite sums over Schur functions, in which the
two sets of flow variables appear on a symmetrical footing. (See \cite{O, 
OS} for further details of such constructions, and additional applications.)

\Subtitle {1a. Schur function expansion of $\tau$ functions}
\smallskip

Suppose we are given:
\hfill
\break
\noindent (i)  A sequence $\{r(m)\}_{m\in{\bf N}}$ of complex numbers.  \hfill
\break \noindent
(ii) A partition
$\l =(n_1\ge \dots \ge n_p\ge 0)$ of $n=\sum_{i=1}^p n_i$ \ . \hfill
\break 
\noindent 
Define, for each integer $M$, the quantity
$$
r_\l(M) := \prod_{i, j} r(M+j-i),  \qquad \forall i,j \in \l   \ ,
\eq rlambda..
$$
where the product is over all pairs $(i,j)$ that lie within a cell in
the corresponding Young diagram (labelled like the entries of a matrix).

  Using the quantities $r_\l(M)$ as coefficients,  we may define an infinite
sum over partitions (without, for the moment, considering the question of
domain of convergence):
$$
\tau_r(M, {\bf t}, {\bf t^*})
:= \sum_\l r_\l(M) s_\l({\bf t}) s_\l({\bf t^*}) \ ,
\eq tauM..
$$
where ${\bf t} = ( t_1, t_2 , \dots )$ and ${\bf t^*} = ( t^*_1, t^*_2 ,
\dots )$ are two infinite sequences of flow variables, and the Schur
function
$s_\l({\bf t})$ is defined by
$$
s_\l({\bf t}) = \det (h_{n_i +j -i}({\bf t})) \eq Schurdet..  
$$
in terms of the elementary Schur functions (complete symmetric functions)
$\{h_j({\bf t})\}_{j\in \bfN}$, which are determined by:
$$
\exp(\sum_{m=0}^\infty t_m z^m)  =\sum_{m=1}^\infty z^m h_m({\bf t}). 
\eq elemSchur..
$$
Whenever the sum \tauM is convergent, it may be shown to be a double
KP $\tau$--function \cite{O, OS}. An indication of how this is done, using
the fermionic Fock space construction, is given in the next subsection.

\Subtitle {1b. Free fermion construction of the $\tau$ function}
\smallskip

 Introduce, on a suitably defined fermionic Fock space, the free fermion
creation and annihilation operators $\{\psi_j, \psi_j^*\}_{j\in {\bf
Z}.} $, which satisfy the usual anti-commutation relations
$$
\eqalign{ \quad [\psi_j, \psi_k]_+&=0, 
\quad 
[\psi_j^*, \psi_k^*]_+ = 0, \cr
[\psi_j, \psi_k^*]_+&= \d_{jk}, \qquad j,k\in {\bf Z}.} \eq..
$$ 
Then for any given sequence $\{r(m)\}_{m\in \{bf N}$ we have the following
fermionic formula for the associated $\tau$-function:
$$
\tau_r(M, {\bf t}, {\bf t^*}) = <M| e^{H({\bf t})} e^{- A({\bf t^*})}|M>, 
\eq..
$$
where
$$
\eqalignno{
H({\bf t}) &:= \sum_{m=1}^\infty t_n H_n, \qquad \qquad
A({\bf t^*}): = \sum_{m=1}^\infty t^*_n A_n, \eqn.a. \cr 
H_n &:= \sum_{k= -\infty}^{\infty} \psi_k  \psi^*_{k+n} , \qquad n\ne 0, 
\eqn.b. \cr 
A_n &:= \sum_{k= -\infty}^{\infty} r(k)  r(k-1) \dots
r(k-n+1)  \psi^*_{k-n} \psi_k ,\eqn.c. \cr }
$$
and $|M>$ is the charge $M$ vacuum state defined by
$$
\eqalign{
|M>&:=\psi_{M-1}\cdots\psi_1\psi_0|0> \ ,  \quad M>0,  \cr
|M>&:=\psi^{*}_{M}\cdots\psi^{*}_{-2}\psi^{*}_{-1}|0> \ ,\quad M<0. } \eq..
$$

Here, the operator $ e^{- A({\bf t^*})}$ may be seen as preparing the
initial state, for fixed values of the parameters ${\bf t^*}$,  while the
operators $ e^{-H({\bf t})}$ generate the abelian group action determinig
the KP flows. (All the $H_n$'s commute amongst themselves, as do the
$A_n$'s.) In this sense, the parameters $\bf t^*$ are just viewed as
distinguishing the elements of an infinite family of KP $\tau$--functions
in the ${\bf t}$ variables. However, the r\^oles of the two sets of parameters
${\bf t}$ and ${\bf t^*}$ may be interchanged and $\tau_r(M, {\bf t}, {\bf
t^*})$ may also be interpreted as a KP $\tau$-function in the ${\bf t^*}$
flow variables for fixed values of
${\bf t}$, thereby defining a double KP $\tau$-function. Moreover, the whole
sequence of $\tau_r(M, {\bf t}, {\bf t^*})$'s for successive values of $M$ may
be shown to define solutions of the $2$-dimensional Toda
hierarchy  \cite{O, OS}.

In what follows, we restrict to the particular choice
$$
 r(n)=n \ ,   \eq..
$$
which gives
$$
r_\l (M) = (M)_\l := \prod_{j=1}^p (M-j)_{n_j} \ ,  \eq..
$$ 
where
$$
(N)_m:= N (N+1) \cdots (N+m-1) \ , \eq..
$$
is the Pochhammer symbol, and denote the corresponding $\tau$--function
$$
\tau_{\rm Id}(M, {\bf t}, {\bf t^*})=:\bfF^M_\l (\bft, \bftt) = \sum_\l
(M)_\l s_\l(\bft) s_\l (\bftt) \ . \eq tauFM..
$$

It is easy to see that for most values of the parameters ${\bf t}$
this is actually a divergent sum, and hence needs some interpretation. This
will be given in the next section. For the present, we just note that there do
exist finite dimensional subregions in the parameter space in which the
sum converges; in fact, there are regions where only a finite number
of terms do not vanish, and hence $\bfF^M_\l (\bft, \bftt)$  is actually
a polynomial in each of the flow variables. Namely, if we choose one of the 
infinite sets of deformation parameters, say ${\bf t}$, to be given in
terms a finite set of  quantities $\{x_1, \dots x_P\}$ by:
$$
t_j := -{1\over j}\sum_{m=1}^P x_m^j, \eq txjnrg..
$$
the elementary Schur functions $h_m({\bf t})$ vanish identically
for $m>P$. Therefore the sums \tauFM  become finite, and $\bfF^M_\l (\bft,
\bftt)$ is just a polynomial in the parameters $(t_1, t_2, \dots)$ and
$(t_1^*, t_2^*, \dots)$.

For $M=1$,  \tauFM takes  a particularly simple form,  since we then have
$r_{(n)}(1)=n!$, and the only partition entering is $\l=(n)$, so 
 \tauFM reduces to the expression
$$
\bfF^1({\bf t},{\bf t^*}) = \sum_{n=0}^\infty n! h_n({\bf t})
h_n({\bf t^*}) \ . \eq tauFone..
$$

\Subtitle {1c. Determinant expression for $\bfF^M({\bf t},{\bf t^*})$ in
terms of  $\bfF^1({\bf t},{\bf t^*})$ }
\smallskip

It is possible to deduce a simple  expression for $\bfF^M({\bf
t},{\bf t^*})$ as a finite determinant in terms of $\bfF^1({\bf t},{\bf
t^*})$ and its derivatives.
\hfill
\break 

\noindent
 {\bf  Proposition 1.1.} {\it  In any region where the sum \tauFM is
uniformly convergent with respect to the parameters $t_1$ and $t_1^*$
the $\tau$-function $\bfF^M({\bf t}, {\bf t^*})$ may be expressed in terms of
$\bfF^1(\bft, \bftt)$ through the following formula:}
$$
  \bfF^M({\bf t}, {\bf t^*})= {1 \over \prod_{j=1}^{M-1} j!}\det
\left({\di^{a+b} \bfF^1({\bf t}, {\bf t^*})\over \di t_1^a \di
t_1^{*b}}\right)_{\{a,b=0, \cdots M-1\}}  \ . \eq FMdet..
$$

This result, and another one similar to it, discussed in the next section,
may be obtained as a consequence of the following well-known lemma:
\hfill
\break

\noindent
{\bf Lemma 1.2:}  {\it For any measure $d\mu(x,y)$ (real or complex) and
any two sets of functions $\{\phi_i(x), \psi_i(y)\}_{i=1, \dots M}$, if the
the following integrals exist along suitably chosen contours $\G_x$ and
${\wt\G}_y$ in the complex $x$- and $y$-planes, they satisfy the identity}
$$
\eqalign{
\int_{\G_x} \int_{{\wt\G}_y} &d\m(x_1,y_1) \dots \int_{\G_x} \int_{{\wt\G}_y}
d\m(x_M,y_M)
\det (\phi_i(x_j))\det (\psi_k(y_l)) \cr
&= M! \det\left(\int_{\G_x} \int_{{\wt\G}_y} d\mu(x,y)
\phi_i(x)\psi_j(y) \right).}  \eq..
$$
The validity of eq. \FMdet may also be shown for more general parameter
values, either as equalities between formal series or in the sense of
asymptotic expansions about a Gaussian point \hbox{$(t_2 \neq 0$}, \ $t_j=0 ,
\ \forall j>2)$.

\noindent
{\bf Proof of Proposition 1.1:} Use
$$
{\di^a h_n \over \di t_1^a}= h_{n-a},   \eq..
$$
(where $h_j$ is understood to vanish for negative $j$) to deduce
$$
{\di^{a+b} \bfF^1({\bft}, {\bftt})\over \di t_1^a \di t_1^{*b}}
=\sum_{m=1}^\infty m! h_{m-a}(\bft) h_{m-b}(\bftt) \ ,  \eq..
$$
and apply the discrete measure version of Lemma 1.2 with:
$$
\phi_i(m):= h_{m-i+1}(\bft), \qquad \psi_j(m) := h_{m-j+1} (\bftt) \ ,
\eq..
$$
using \Schurdet.

\goodbreak
%=========================== Section 2. ================================
\section 2. Relation between $\bfF^M({\bf t},{\bf t^*})$ and matrix
integrals
\smallskip \nobreak

\Subtitle {2a. The $2$-matrix partition function ${\bf Z}_{2P}^M$ }
\smallskip
\nobreak

 The unitary $2$-matrix model partition function for $M \times M$ hermitian
matrices \cite{IZ, DKK, MS, BEH}, after integrating out the ``angular''
variables,  is given, up to a proportionality constant, by the following
multiple integral over the eigenvalues.
$$
\bfZ^M_{2P}({\bf t}, {\bf t^*}) :=  \int_{\G_{x_1} \times
\G_{y_1}}\dots \int_{
\G_{x_M}\times \G_{y_M}} \prod_{a=1}^M\Delta({\bf x}) \Delta({\bf
y}) e^{(\sum_{j=1}^{\infty}(t_j x_a^j +t_j^* y_a^j -x_a y_a)} dx_a dy_a  \
, \eq partMtwo..
$$
where
$$
\Delta({\bf x}):= \prod_{i>j =1}^M (x_i - x_j), \qquad
\Delta({\bf y}):= \prod_{i>j =1}^M (y_i - y_j) \ .
$$
(Although for the case of Hermitian matrices, the integration is along the
real axes, we retain here the possibility of integration over more general
contours $\G_x$ and $\wt{\G}_y$ in the $x$ and $y$ planes.)

 This may also be expressed as a determinant in terms of derivatives of 
$\bfZ^M_{2P}({\bf t}, {\bf t^*})$: \hfill \break

\noindent
{\bf  Proposition 2.1}: 
{\it We have the following expression for the $2$-matrix partition function in
terms of the $M=1$ case, valid in any region where the integral \partMtwo is
uniformly convergent with respect to the parameters $t_1$ and $t_1^*$.}
$$
\bfZ^M_{2P}({\bf t}, {\bf t^*})=M! \det \left({\di^{a+b} \bfZ^1_{2P}({\bf t}, 
{\bf t^*})\over \di t_1^a \di t_1^{*b}}\vert_{a,b=0, \cdots M-1}\right) \ .
\eq ZMdet..  
$$
The proof of this result is also a straightforward application of Lemma 1.2.

\Subtitle {2b. Schur function expansion of  $\bfZ^M_{2P}$ }
\smallskip

   We now state the main result, relating the tau function $\bfF^M({\bf t},
{\bf t^*})$ to the $2$-matrix partition function $\bfZ^M_{2P}({\bf t}, {\bf
t^*})$. \hfill \break

\noindent {\bf Theorem 2.2:} {\it In any region where both the integral
\partMtwo defining the partition function $\bfZ^M_{2P}(\bft, \bftt)$ and the
sum \tauFM defining the $\tau$ function $\bfF^M(\bft, \bftt)$ are
uniformly convergent, we have the equality:}
$$
\bfZ^M_{2P}(\bft, \bftt) = 2\pi (\prod_{j=1}^M j!) \bfF^M(\bft, \bftt).  \eq
ZMFM..
$$

 More generally, when the integral \partMtwo defining $\bfZ^M_{2P}(\bft,
\bftt)$ is convergent, whether the sum \tauFM is convergent or not, the
former may be viewed as a multi-variable Borel sum of the series \tauFM
defining $\bfF^M(\bft, \bftt)$. Moreover, even when the integral is not
convergent, it may be  expanded in a perturbation series about a Gaussian
point where $t_2\neq 0$, $t_n=0, \ \forall n>2$, and the resulting asymptotic
series is given by the sum \tauFM.  A  proof of these result is given in
\cite{HO}.  Here, we just give a sketch of how the identification as a Borel
sum may be derived for the case
$M=1$, and conclude, from Props. 1.2 and 2.1  that this implies the analogous
result in the multiple integral case. We also indicate a purely formal
approach directly for arbitrary $M$, which does not address the question of
convergence. 

\Subtitle {2c. Formal expansions for $M=1$}
\smallskip

We first show how the relation \ZMFM may be derived formally, without
considerations of convergence, for the case $M=1$.
\hfill \break

\noindent
{\bf Proposition 2.3:} {\it  Choosing the integration contours $\Gamma_x$ and
${\wt \Gamma}_y$ to be orthogonal lines in the complex plane, the following
identity holds:}
$$
 \bfZ^1_{2P}({\bf t}, {\bf t^*})  = 2\pi i\bfF^1({\bf t}, {\bf t^*}) \ .  
\eq..
$$
as formal expansions in powers of the variables ${\bf t}= (t_1, t_2, \dots)$
and ${\bf t^*}=(t^*_1, t^*_2, \dots)$.

\noindent
 {\bf Proof:}
Substituting the series \elemSchur generating the elementary  Schur functions
into $\bfZ^1_{2P}$ gives:
$$
\eqalign{
\bfZ^1_{2P}({\bf t}, {\bf t^*})
 &= \int_{\G_x}dx\int_{\G_y} dy
\sum_{m=1}^\infty x^m h_m({\bf
t})\exp(\sum_{j=1}^{\infty}t^*_j y^j)e^{- x y} \cr
&= \int_{\G_x}dx\int_{\G_y } dy
\sum_{m=1}^\infty  h_m({\bf
t})\exp(\sum_{j=1}^{\infty}t^*_j y^j)(-1)^m{d^m\over d y^m} e^{-x y}
} \eq..
$$
Taking the derivatives with respect to $y$ outside the $x$ integral, and
evaluating the $x$ integral along the imaginary axis, we obtain a
$\d$-function, giving 
$$
\eqalign{
\bfZ^1_{2P}({\bf t}, {\bf t^*})  &= 2\pi i\int_{\G_y } d y
\sum_{m=1}^\infty (-1)^m h_m({\bf t}){d^m\over
d y^m}\d(y)\exp(\sum_{j=1}^{\infty}t^*_j y^j) \cr
&=2\pi i \sum_{m=1}^\infty  h_m({\bf t}){d^m\over
d y^m} \left(\sum_{j=1}^{\infty}h_j({\bf t^*}) y^j\right)_{y=0} \cr
&=
2\pi i \sum_{m=1}^\infty m!  h_m({\bf t})h_m({\bf t^*})= 
2\pi  i\bfF^1({\bf t}, {\bf t^*}) \ .} \eq..
$$
\smallskip \noindent

\Subtitle {2d. Borel sum intepretation of eq. \ZMFM for $M=1$}
\smallskip

We now show, for the case $M=1$, how $\bfZ^1_{2P}({\bf t}, {\bf t^*})$ may be
interpreted as a Borel sum for \tauFone even when the sum diverges.

To do this, we reinterpret the integrals defining $\bfZ^1_{2P}({\bf t}, {\bf
t^*})$ in such a way that the variables $x$ and $y$ are replace instead by
a pair of complex conjugate variables $z$ and ${\bar z}$ in the complex
plane, with the integral viewed simply as a double integral in the plane.

We then have, using the expansion \elemSchur :
$$
\eqalignno{
\int \int  dz  d{\bar z} 
e^{\sum_{j=1}^\infty (t_j z^j   t^*_j {\bar z}^j  -z {\bar z}} 
&= \int_0^\infty r dr \int_0^{2\pi} d\phi 
\sum_{j=1}^\infty r^j h_j({\bf t}) 
\sum_{k=1}^\infty r^k  h_k({\bf t^*}) e^{i(k-j)\phi}e^{-r^2} \cr
&=2 \pi \int_0^{\infty} r dr \sum_{j=1}^\infty 
 r^{2j} h_j({\bf t})  h_j({\bf t^*})  e^{-r^2} \cr
&= \pi \int_0^{\infty} dx \sum_{j=1}^\infty 
x^j h_j({\bf t})  h_j({\bf t^*})  e^{-x}
\eqn..}
$$
and this is just the Borel sum formula for the series \tauFone.

\Subtitle {2e.  Sketch of a formal proof of Theorem 2.2 using the
generating expansion}
\smallskip

The following identity \cite{Ma} gives a generating function expansion for
Schur functions:
$$
\exp({\sum_{n=0}^\infty n t_n t_n^*}) = \sum_\l s_\l (\bft) s_\l (\bftt), 
\eq..
$$
 Applying this identity to the cases where:
 $$
\eqalignno{\bftt &=[\bfx] := (\sum_{i=1}^{\infty} x_i, {1\over
2}\sum_{i=1}^{\infty} x_i^2,
\dots),
\qquad
\bfx =(x_1, \dots x_M) \ , \eqn.a.\cr
 \bft &=[\bfy]  :=(\sum_{i=1}^{\infty} y_i, {1\over
2}\sum_{i=1}^{\infty} y_i^2, \dots), \qquad \bfy =(y_1, \dots y_M)  \ ,
\eqn.b.}
$$ 
gives the expansions
$$
\eqalignno{\exp(\sum_{j=1}^M \sum_{m=0}^\infty t_m x_i^m) &= \sum_\l
s_\l(\bft)s_\l([\bfx]), \eqn.a.\cr
 \exp(\sum_{j=1}^M \sum_{m=0}^\infty t^*_m y_j^m)
&= \sum_\mu s_\mu(\bftt)s_\mu([\bfy]). \eqn .b.}
$$
\nextnumber
Substituting these into the expression for the $2$-matrix partition
function, using the relations 
$$
s_\l ([\bfx])\Delta(\bfx) = \det(x_i^{n_j-j+M}), \qquad s_\l
([\bfy])\Delta(\bfy) = \det(y_k^{m_l-l+M}) \ , \eq..
$$ 
and once again applying Lemma 2.2, the integrals may be evaluated along
suitably chosen contours (as in the formal proof above for $M=1$), to
obtain the stated result:
$$
\bfZ^M_{2P}(\bft, \bftt) = \prod_{j=0}^M j! \sum_\l (M)_\l s_\l (\bft)
s_\l(\bftt) \ .  \eq..
$$

   The detailed proof, and a further discussion of the interpretation of the
sum $\tauFM$ as an asymptotic series about the Gaussian point may be found in
\cite{HO}.

 %%%%%%%%%%%%%%%%%%%%%%%%%%%%%%% Acknowledgements %%%%%%%%%%%%%%%%%%%%%%%%%%%
\bigskip\bigskip \noindent{\it Acknowledgements.}
The authors thank M. Bertola and B. Eynard  for helpful discussions. This
research was supported in part by the Natural Sciences and Engineering
Research Council of Canada.
\bigskip \bigskip
%%%%%%%%%%%%%%%%%%%%%%%%%%%%%%% References %%%%%%%%%%%%%%%%%%%%%%%%%%%%%%%%%
  \centerline{\bf References} \hfill 
 {\smalltype
\item{\bf [BEH]} M. Bertola, B. Eynard and  J. Harnad,  ``Duality,
Biorthogonal Polynomials and Multi-Matrix Models'',{\it  Commun. Math. Phys.}
{\bf 229}, 57-71 (2002), nlin.SI/0108049.
\item{\bf [DKK]} J.M. Daul, V. Kazakov, I.K. Kostov, ``Rational Theories of 
2D Gravity from the Two-Matrix Model'', {\it Nucl. Phys.} {\bf B409}, 311-338 
(1993), hep-th/9303093.
\item{\bf [HO]} J. Harnad and A. Yu.  Orlov  ``Schur function
expansions of matrix integrals'', preprint CRM (2002), in preparation.
\item{\bf [IZ]} C. Itzykson and J.B. Zuber, ``The planar approximation 
(II)'', {\it J. Math. Phys.} {\bf 21}, 411 (1980).
\item{\bf [MS]} M.L. Mehta and P. Shukla, ``Two coupled matrices:
Eigenvalue correlations and spacing functions'', {\it J. Phys. A: Math. Gen.}
{\bf 27}, 7793-7803 (1994).
\item{\bf [Ma]} Macdonald, M.~L., {\it Symmetric polynoimials}, 2nd edition
\item{\bf [O]} Orlov, A. Yu. ``Soliton theory, symmetric function
and matrix integrals'' nlin.SI/0207030 (2002).
\item{\bf [OS]} Orlov, A. Yu. and Scherbin, M.D., ``Milne's hypergeometric
functions in terms of free fermions''. {\it J. Phys. } {\bf A} Math. Gen. 
{\bf 34} 2295-2310, (2001).

}
\vfill \eject

\end